\documentclass[twocolumn]{aastex631}
\usepackage{mathrsfs}
\usepackage{amsmath}
\usepackage{enumitem}
\usepackage[T1]{fontenc}

\newcommand{\ergs}{\rm erg\,s^{-1}}
\newcommand{\gcm}{\rm g\,cm^{-3}}
\newcommand{\kms}{\rm km\,s^{-1}}

\newcommand{\Msun}{M_{\sun}}

\newcommand{\sigT}{\sigma_{\rm T}}

\newcommand{\sigv}{\sigma_{\rm v}}

\newcommand{\dotM}{\dot{m}}

\newcommand{\vterm}{v_{\rm term}}

\newcolumntype{x}{>{\color{blue}}c}

\def\ihep{State Key Laboratory of Particle Astrophysics, Institute of High Energy Physics,
Chinese Academy of Sciences, 19B Yuquan Road, Beijing 100049, China}

\def\KIAA{Kavli Institute for Astronomy and Astrophysics, Peking University, Beijing 100871, People’s Republic of China}

\def\PKU{Department of Astronomy, School of Physics, Peking University, Beijing 100871, People’s Republic of China}

\begin{document}

\title{Optically Thick Outflow Driven by Supercritical Accretion May Explain Little Red Dots}

\author[0000-0003-3086-7804]{Jun-Rong Liu}
\affil{\ihep}

\author[0000-0001-7584-6236]{Hua Feng}
\affil{\ihep}
\correspondingauthor{Hua Feng}
\email{hfeng@ihep.ac.cn}

\author[0000-0001-6947-5846]{Luis C. Ho}
\affil{\KIAA}
\affil{\PKU}


\begin{abstract}
Recent JWST observations have revealed a population of compact, optically red sources known as Little Red Dots (LRDs).
A popular interpretation is that LRDs host massive black holes embedded in dense gaseous envelopes, yet the physical origin of such envelopes remains unclear.
We propose that the optically thick outflow driven by supercritical accretion onto black holes may explain the envelope.
Based on an analytic radiative hydrodynamic outflow model, we relate the outflow properties to the black hole mass $M$ and dimensionless accretion rate $\dotM$.
With $M\sim10^5-10^7\,\Msun$ and $\dotM\sim 1500-5000$, the outflow photosphere reaches luminosities of $10^{43}-10^{45}\,\ergs$ and temperatures of $\sim 3000-6000$ K, effectively matching the red optical continua observed in LRDs.
The presence of a thick scattering region beyond the photosphere is central to deciphering the distinctive properties of LRDs.
For each object, if one derives $M$ and $\dotM$ from the observed luminosity and temperature, while accounting for both kinematic broadening and scattering effects, the model predicts an emission line FWHM consistent with observations within a factor of 1.5 for more than 80\% objects. 
Furthermore, with Cloudy simulations, we find that the partially ionized gas beyond the photosphere produces Balmer breaks
broadly consistent with measurements.
\end{abstract}


\section{Introduction}

In recent surveys with the James Webb Space Telescope (JWST), a population of compact and extremely red sources, termed Little Red Dots (LRDs), has been identified \citep{Labbe2023, Greene2024, Kokorev2024, Matthee2024, Kocevski2025, Labbe2025}.
These objects are typically unresolved in JWST imaging, implying physical sizes no larger than a hundred parsecs.
Many show red optical continua together with broad Balmer features and V-shaped spectral shapes \citep{Greene2024, Rusakov2026}.
Their inferred number densities at redshifts $z\gtrsim4$ are high compared with expectations from extrapolated quasar luminosity functions \citep{Matthee2024,Inayoshi2024}, and may require new channels of black hole growth \citep{Hu2006, Jiang2026}.
Their weak X-ray \citep{Yue2024,Kocevski2025} and radio emission \citep{Gloudemans2025,Perger2025,Mazzolari2026} challenges the simple interpretation of standard unobscured active galactic nuclei (AGNs).

The physical nature of LRDs remains poorly understood, prompting a wide range of theoretical explanations, including dense gas envelopes \citep{Inayoshi2025,Kido2025}, dusty outflows \citep{Labbe2025,Li2025}, supermassive stars \citep{Zwick2025,Nandal2026,Chisholm2026}, late-stage quasi-stars \citep{Santarelli2025,Begelman2026}, accretion-modified stellar populations in AGN disks \citep{Wang2025}, outer AGN disks \citep{Zhang2026,Chen2026b}, direct-collapse black holes \citep{Jeon2026}, and purely stellar origin \citep{Wang2024,Akins2025}.
Notably, LRDs may represent a diverse population \citep[e.g., see review in][]{Inayoshi2025c}.
Among these, one popular interpretation invokes a dense envelope surrounding an accreting SMBH, referred to as a ``black hole star" \citep{Kido2025,Graaff2025,Umeda2026,Rusakov2026,Sneppen2026,Sun2026}.
In this picture, a photosphere with a temperature of order $5\times10^3$ K explains the red optical continuum \citep{Kido2025,Graaff2025,Umeda2026}.
The same framework can also account for Balmer breaks, colors, and absorption features \citep{Liu2025,Liu2026},
while the observed exponential broad-line wings may arise from multiple electron scattering \citep{Eugenio2025,Rusakov2026,Sneppen2026}.
Dense gas may also strongly absorb the intrinsic X-ray and radio emission from the central source \citep{Maiolino2025,Trinca2026}.

Radiation-driven outflows are an inevitable outcome of supercritical accretion at the inner accretion flow \citep{Shakura1973}.
It has been argued that such outflows should be massive and optically thick \citep{Meier1982, King2003, Kitaki2021}, manifesting themselves as a thermal component.
They have been invoked to explain the soft excess ($\sim0.1 - 0.4$\,keV) seen in supercritically accreting stellar compact objects such as ultraluminous X-ray sources \citep{Poutanen2007, Shen2015, Shen2016, Zhou2019, Yao2019, Qiu2021} and tidal disruption events with relatively low temperatures \citep[$\sim 10^4~$K,][]{Loeb1997,Miller2015,Miller2015b,Dai2018}.

In this work, we apply the analytic radiation-driven outflow model of \citet{Meier1982} to LRDs and explore whether the observed envelope can be interpreted as the photosphere of an optically thick outflow.
We try to self-consistently interpret the observed red continuum (its blackbody temperature and luminosity), broad-line FWHMs, and the Balmer breaks and decrements within this framework.

The paper is organized as follows.
In Section \ref{sect:model}, we briefly review the model.
In Section \ref{sect:LRD}, we try to explain the observed properties of LRDs with the model.
In Section \ref{sect:discuss}, we discuss the self-consistency of the model.
In Section \ref{sect:conclusion}, we summarize the main results.

\begin{figure}
\centering
\includegraphics[width=\linewidth,trim=0 0 0 0]{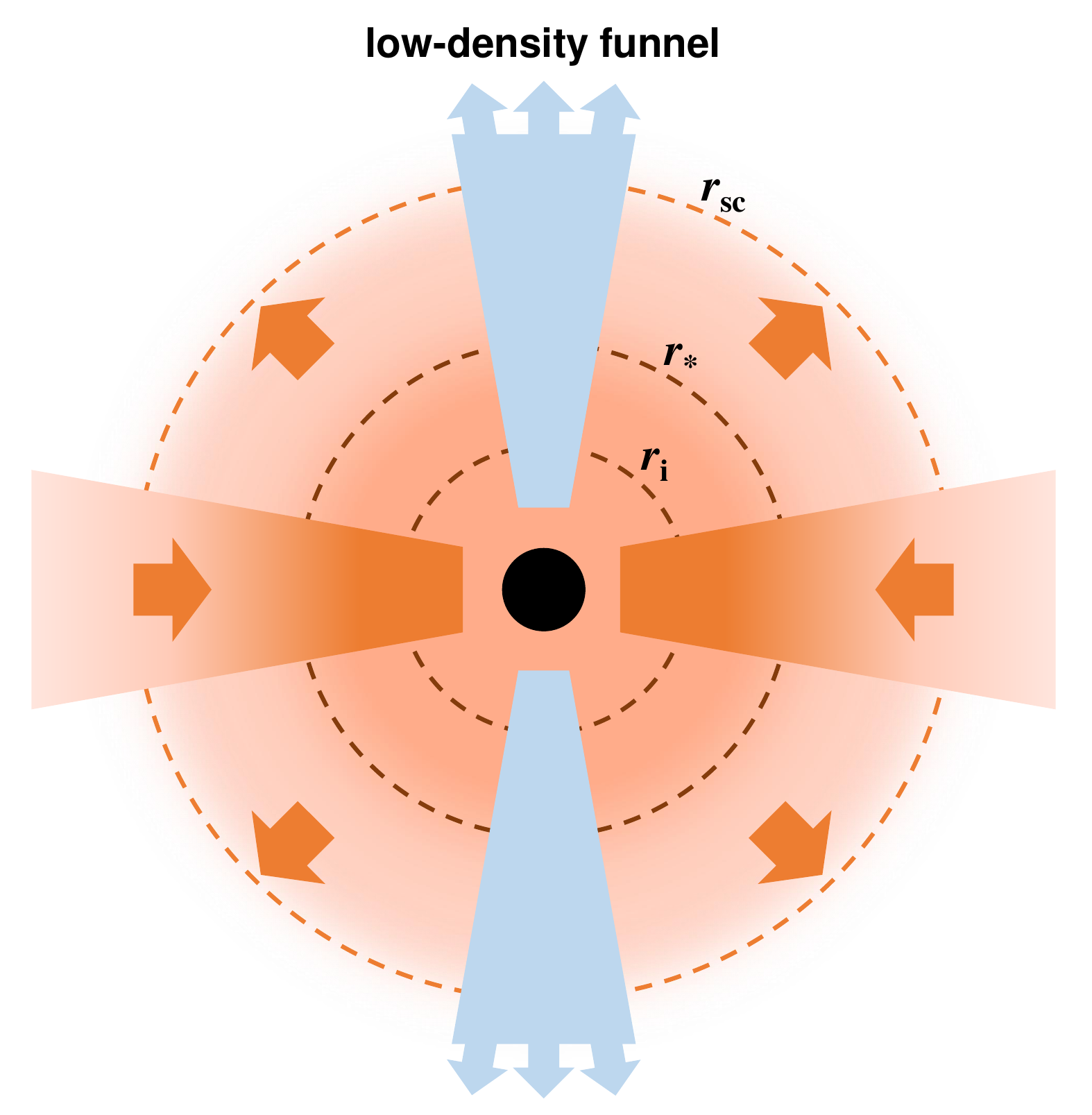}
\caption{Schematic illustration of the outflow model.
The arrows indicate the inflow and outflow. 
The outflow starts from $r_{\rm i}$ and then undergoes acceleration and free expansion. 
$r_\ast$ denotes the photosphere and $r_{\rm sc}$ denotes the scattersphere.
UV radiation from the central accretion flow can escape from the low-density funnels (blue cones).
}
\label{fig:wind}
\end{figure}

\section{Model}
\label{sect:model}


\citet{Meier1982} found analytic solutions of a spherically asymmetric, radiation-driven, and optically thick outflow. 
\citet{Meier2012} further connected the outflow with a slim disk. 
Readers can refer to the Appendix of \citet{Zhou2019} for a concise but complete description of this particular model; we use the same set of terminology and definitions in this work. 

Here, we briefly review the assumptions, logic, and results of the model without delving into detailed equations.
There are two fundamental parameters, the black hole mass $M$ and the dimensionless accretion rate $\dotM$ ($\dotM = 1$ corresponds to an accretion rate just sufficient to power the Eddington limit).

The model assumes that completely coupled plasmas and radiation are launched at the injection radius ($r_{\rm i}$) and then develop outwards. 
The injection radius is assumed to be the trapping radius ($r_{\rm trap}=6GM\dot{m}/c^2$ for a non-spinning black hole, where $G$ is the gravitational constant and $c$ is the speed of light) of the standard slim disk \citep{Abramowicz1988}, where advective and radiative cooling become comparable and therefore there is almost no photon trapping in the outflow.
At $r_{\rm i}$, the radiation pressure removes excessive accretion materials into the outflow, resulting in an inflow with $\dot m \approx 1$ to continue powering the system.
The outflow first goes through acceleration until the adiabatic radius ($r_{\rm ad}$) and then follows free expansion. 
From the observational point of view, there are two characteristic radii of interest, the photosphere of the outflow, $r_\ast$, where the effective absorption optical depth equals unity, and the last scattering radius, $r_{\rm sc}$, also termed as the scattersphere, beyond which the radiation decouples from matter. 
A schematic drawing of the model is shown in Figure~\ref{fig:wind}. 

As local thermodynamic equilibrium is always assumed within the photosphere, a natural consequence is that the emergent luminosity (that observed by an observer at infinity) by definition equals the input luminosity at $r_i$, which is approximately the Eddington luminosity.
In the specific slim disk adopted here, the observed blackbody luminosity is $L_{\rm bb}=3L_{\rm Edd}/4$, following Eq.\ (14) of \cite{Zhou2019}.

Therefore, one expects to observe a blackbody-like spectrum from the outflow.
Here we rewrite the equations that are used in this work, including the observed blackbody temperature (equals the photospheric temperature, $T_{\rm bb} = T_\ast$), blackbody luminosity, terminal velocity, photospheric radius, and density at the photosphere as follows:
\begin{align}
\label{eq:temp}
T_{\rm bb} &= 5.80 \times 10^{3} \, M_6^{-2/11} \left(\frac{\dotM_3}{2}\right)^{-15/11} \, \rm K , \\
\label{eq:lum}
L_{\rm bb} &= 1.09 \times 10^{44} \, M_6 \, \ergs ,\\
\label{eq:vel}
\vterm &= 7.46 \times 10^2 \, \left(\frac{\dotM_3}{2}\right)^{-1/2} \, \kms,\\
\label{eq:rstar}
r_\ast &= 5.22 \times10^{16} \, M_6^{10/11} \left(\frac{\dotM_3}{2}\right)^{51/22} \, \rm cm,\\
\label{eq:rho}
\rho_\ast &= 1.51 \times10^{-15} \, M_6^{-9/11} \left(\frac{\dotM_3}{2}\right)^{-69/22} \gcm ,
\end{align}
where $M_6 = M / (10^6 M_\sun)$ and $\dot m_3 = \dot m / 10^3$.
We note that $T_\ast$ has another solution in the case when $r_{\rm ad} > r_\ast$, but this never occurs in the parameter space of this work.
 
Other parameters include the accretion efficiency assumed to be 0.1, accretion viscosity $\alpha$ assumed to be 0.1, metallicity assumed to be solar abundance, and average molecular weight assuming full ionization; all these parameters have minor influences to the results.
We note that in the early Universe, a low metallicity is generally expected and affects the opacity in this model. 
We tried with 0.1 times solar abundance and found that the electron-scattering opacity varies from $\kappa_{\rm es}\simeq0.340$ to $\simeq0.344~{\rm cm^2~g^{-1}}$ in the assumption of full ionization \citep{Kippenhahn2013}, and therefore has a negligible effect.

\begin{deluxetable*}{lcccc ccccc c}
\footnotesize
\tablecaption{Measured properties of LRDs and comparison with the outflow model.}
\label{tab:LRD}
\colnumbers
\tablehead{
Object & $z$ & $T_{\rm bb}$ & $\log [L_{\rm bb} / (\ergs)]$ & FWHM$_{\rm obs}$ & $M$ & $\dotM$ & $\vterm$ & $\sigv$ & FWHM$_{\rm mod}$ & Ref.\\
& & (K) & \colhead{} & $(\kms)$ & $(10^6\Msun)$ & & $(\kms)$ & $(\kms)$ & $(\kms)$  }
\startdata
\textit{The Egg} & 0.1 & $4761$ & $43.328$ & $965$ & 0.20 & 2873 & 622 & 848 & 1910 & (1) \\
JADES-28074 & $2.27$ & $5038^{+15}_{-16}$ & $44.484^{+0.001}_{-0.001}$ & $3610 \pm 22$ & $2.80$ & $1933$ & $758$ & $1336$ & $2717$ & (4) \\
GTO-2974 & $2.32$ & $4193^{+116}_{-135}$ & $43.401^{+0.007}_{-0.008}$ & $1420 \pm 178$ & $0.23$ & $3084$ & $600$ & $773$ & $1785$ & (3) \\
UNCOVER-20698 & $2.42$ & $2359^{+13}_{-12}$ & $43.730^{+0.001}_{-0.001}$ & $1707 \pm 810$ & $0.49$ & $4251$ & $511$ & $509$ & $1341$ & (3) \\
NEXUS-20152 & $2.87$ & $5119^{+226}_{-177}$ & $43.946^{+0.011}_{-0.009}$ & $659 \pm 332$ & $0.81$ & $2254$ & $702$ & $1118$ & $2361$ & (5) \\
RUBIES-40579 & $3.11$ & $3648^{+27}_{-26}$ & $44.667^{+0.002}_{-0.002}$ & $2466 \pm 31$ & $4.28$ & $2315$ & $693$ & $1043$ & $2251$ & (2) \\
NEXUS-3713 & $3.22$ & $4205^{+392}_{-246}$ & $43.973^{+0.039}_{-0.026}$ & $946 \pm 219$ & $0.87$ & $2582$ & $656$ & $936$ & $2067$ & (5) \\
RUBIES-144195 & $3.35$ & $5703^{+252}_{-280}$ & $43.836^{+0.010}_{-0.013}$ & $2053 \pm 130$ & $0.63$ & $2154$ & $718$ & $1193$ & $2478$ & (2) \\
RUBIES-154183 & $3.55$ & $3973^{+58}_{-54}$ & $44.194^{+0.004}_{-0.004}$ & $2481 \pm 17$ & $1.44$ & $2515$ & $665$ & $957$ & $2104$ & (2) \\
RUBIES-23438 & $3.69$ & $4217^{+499}_{-462}$ & $43.529^{+0.034}_{-0.029}$ & $2058 \pm 183$ & $0.31$ & $2952$ & $614$ & $809$ & $1849$ & (2) \\
JADES-13329 & $3.94$ & $4248^{+488}_{-330}$ & $43.567^{+0.038}_{-0.026}$ & $1341 \pm 230$ & $0.34$ & $2903$ & $619$ & $825$ & $1875$ & (4) \\
RUBIES-167741 & $4.12$ & $3804^{+1601}_{-607}$ & $43.566^{+0.125}_{-0.079}$ & $2399 \pm 98$ & $0.34$ & $3149$ & $594$ & $745$ & $1743$ & (2) \\
RUBIES-31747 & $4.13$ & $3826^{+260}_{-243}$ & $43.753^{+0.018}_{-0.015}$ & $2277 \pm 150$ & $0.52$ & $2961$ & $613$ & $796$ & $1830$ & (2) \\
JADES-73488 & $4.13$ & $6380^{+405}_{-336}$ & $43.878^{+0.033}_{-0.027}$ & $2228 \pm 48$ & $0.69$ & $1958$ & $754$ & $1349$ & $2728$ & (4) \\
RUBIES-119957 & $4.15$ & $5343^{+2032}_{-672}$ & $43.732^{+0.142}_{-0.063}$ & $1977 \pm 89$ & $0.50$ & $2332$ & $690$ & $1081$ & $2297$ & (2) \\
CAPERS-19799 & $4.22$ & $3659^{+127}_{-117}$ & $43.862^{+0.008}_{-0.008}$ & $2194 \pm 60$ & $0.67$ & $2958$ & $613$ & $791$ & $1826$ & (2) \\
NEXUS-2314 & $4.42$ & $5303^{+600}_{-437}$ & $44.890^{+0.118}_{-0.101}$ & $2751 \pm 256$ & $7.14$ & $1644$ & $822$ & $1639$ & $3197$ & (5) \\
UNCOVER-45924 & $4.46$ & $5708^{+15}_{-20}$ & $44.917^{+0.001}_{-0.001}$ & $4540 \pm 50$ & $7.60$ & $1544$ & $848$ & $1784$ & $3421$ & (6) \\
\enddata
\tablecomments{
Column (1): source name.
Column (2): redshift.
Column (3): observed temperature of the red envelope. 
Column (4): observed luminosity of the red envelope. 
Column (5): observed FWHM for the broad-line component fitted with a Gaussian function.
Column (6): best-fit black hole mass given $T_{\rm bb}$ and $L_{\rm bb}$.
Column (7): best-fit accretion rate given $T_{\rm bb}$ and $L_{\rm bb}$. 
Column (8): terminal velocity given $M$ and $\dotM$.
Column (9): characteristic broadening scale given $M$ and $\dotM$.
Column (10): model predicted FWHM given $M$ and $\dotM$.
Column (11): reference. 
}
\tablerefs{
(1) \citet{Lin2026}, 
(2) \citet{Hviding2025}, 
(3) \citet{Wang2026}, 
(4) \citet{Juodzbalis2026}, 
(5) \citet{Zhuang2026}, and
(6) \citet{Labbe2024}.}
\end{deluxetable*}

\begin{figure}
\centering
\includegraphics[width=\linewidth,trim=5 5 0 0]{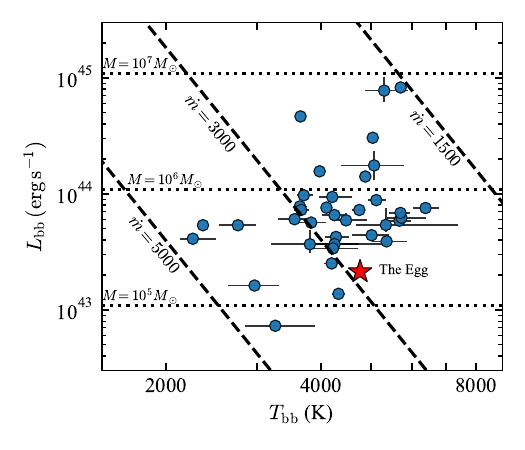}
\caption{Blackbody luminosity vs.\ temperature measured from the red envelope in LRDs (circles) and the local analog ``\textit{The Egg}'' (star).
The dashed and dotted lines show the predicted blackbody luminosity and temperature from the outflow model at different black hole masses ($M = 10^5$, $10^6$, and $10^7 \Msun$) and accretion rates ($\dotM = 1500$, $3000$, and $5000$).}
\label{fig:bb}
\end{figure}

\begin{figure}
\centering
\includegraphics[width=\linewidth,trim=5 5 0 0]{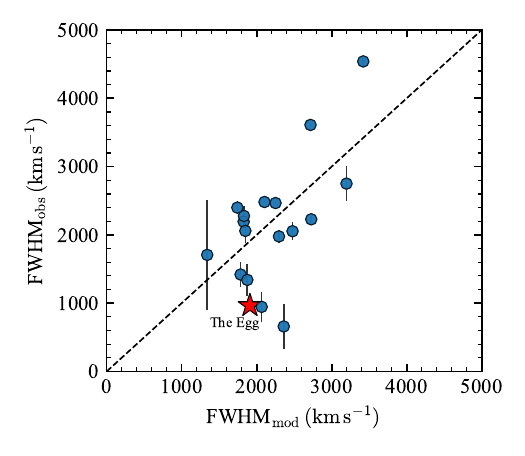}
\caption{Observed vs.\ model predicted FWHMs of the broad-line component. Both kinetic and scattering effects are taken into account when calculating the velocity dispersion. The dashed line marks the 1:1 relation.
}
\label{fig:fwhm}
\end{figure}

\section{temperatures, luminosities, line widths, and Balmer features}
\label{sect:LRD}

In this section, we try to investigate if the outflow model can account for the envelope or dense cocoon seen in LRDs \citep{Kido2025,Umeda2026}.
We selected 36 LRDs whose spectra have been successfully fitted with a modified blackbody model \citep{Graaff2025}.
We verified that the modified blackbody has a temperature and luminosity close to those of a pure blackbody, with an accuracy sufficient for our purpose.
We also included the local LRD analog known as \textit{The Egg} \citep{Lin2026}.
The observed blackbody luminosity and temperature span $10^{43}-10^{45}\,\ergs$ and $2000-7000~$K, respectively, shown in Figure~\ref{fig:bb}.
We plotted the theoretical blackbody luminosity and temperature given the outflow model with different $M$ and $\dotM$ in the same figure with Eqs.~(\ref{eq:temp}-\ref{eq:lum}).
As one can see, to match the observed ranges of luminosity and temperature, one requires $10^5 \Msun < M < 10^7 \Msun$ and $1500 < \dotM < 5000$.

At the photosphere, given typical values of $n_{\rm H}=5\times 10^8\,\rm cm^{-3}$ and $T_{\rm bb}=5000$ K, we estimated an ionization fraction of $x=0.2$ using the Saha equation
\begin{equation}
\frac{x^2}{1-x}=\left(\frac{2\pi m_{\rm e}k_{\rm B}T_{\rm bb}}{h^2}\right)^{3/2}\frac{e^{-I_0/k_{\rm B}T_{\rm bb}}}{n_{\rm H}} \; ,
\end{equation}
where $m_{\rm e}$ is the electron mass,
$k_{\rm B}$ is the Boltzmann constant,
$h$ is the Planck constant,
and $I_0=13.6$ eV is the ground-state ionization energy of hydrogen.
This estimate indicates that the gas near the photosphere is partially ionized, thereby contributing to the observed Balmer features.
Deeper inside the photosphere, where the temperature and radiation energy density are significantly higher, the gas is expected to be highly or fully ionized, and the model scalings that rely on 
the ionized-gas approximation remain valid.

Furthermore, to examine if the outflow model can reproduce the observed broad-line velocity dispersion, we collected the broad-line widths measured for 18 of them from the literature, where the broad-line component was fitted with a Gaussian function and the FWHM is listed in Table~\ref{tab:LRD}.
For each source, we derived the best-fit $M$ and $\dotM$ given the measured blackbody luminosity and temperature, and then calculated the terminal velocity $v_{\rm term}$ using Eq.~(\ref{eq:vel}), also listed in Table~\ref{tab:LRD}. 
We note that in this study the photosphere is always located in the free-expansion region, suggesting that the outflow velocity at $r_\ast$ is already $v_{\rm term}$.
We assume that the kinematic broadening produces a top-hat profile with a width of $2 v_{\rm term}$.
\begin{equation}
F_0(v)=
\begin{cases}
\displaystyle \frac{1}{2v_{\rm term}}, & |v|\le v_{\rm term},\\
0, & |v|>v_{\rm term}.
\end{cases}
\end{equation}

Radiative transfer in the outflow further modifies the emergent line profile in addition to the kinematic velocity dispersion \citep[e.g.,][]{Chang2026}.
In particular, Thomson scattering can broaden the line profiles \citep{Weymann1970,Laor2006} and produce exponential wings.
Such a feature has been seen in the broad-line component of LRDs \citep{Rusakov2026,Matthee2026,Ji2026}.
We adopted an exponential kernel $\mathcal{E}(\Delta v)=\exp({-|\Delta v|}/\sigv)/2\sigv$
derived from Monte Carlo simulations \citep{Rusakov2026},
where $\Delta v=(\lambda-\lambda_0)c/\lambda_0$,
$\lambda$ is the wavelength,
$\lambda_0$ is the line center,
$\sigv=(428\tau_{\rm es}+370)\sqrt{T_{\rm bb}/(10^4\rm\,K)}\rm\,\kms$ is the characteristic broadening scale,
and $\tau_{\rm es}$ is the electron scattering optical depth.
Because the ionization structure is not solved self-consistently in the analytic model, we adopted an ionization fraction of $x=0.1$ as a fiducial value for estimating the scattering contribution.
We then obtained $\tau_{\rm es}=xn_{\rm H} r_\ast \sigT$ for each LRD.
The FWHM of exponential wing is $2\sigv\ln 2$, listed in Table~\ref{tab:LRD}.
We convolved the top-hat intrinsic velocity profile with the exponential electron-scattering kernel
\begin{equation}
F_{\rm mod}(v)=\int_{-\infty}^{+\infty}
F_0(v')\mathcal{E}(v-v')\,{\rm d}v' \; ,
\end{equation}
and derived the FWHM of emergent emission line (see Table \ref{tab:LRD}).
We plotted the measured FWHM vs.\ model predicted FWHM in Figure~\ref{fig:fwhm}.
As one can see, they are broadly consistent with each other, with 15 out of 18 objects differing by less than a factor of 1.5.
We also note that, the model line profile has a Gaussian-like core plus exponential wings, well consistent with some observations \citep{Rusakov2026,Matthee2026}.

\begin{figure}
\centering
\includegraphics[width=\linewidth]{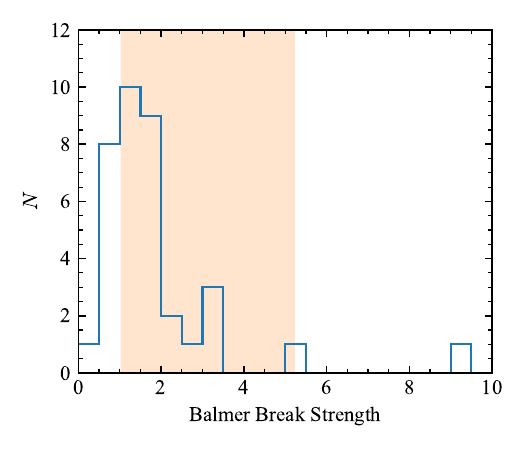}
\caption{Distribution of Balmer break strengths in LRDs.
The shaded region shows the expected range from the fiducial outflow model with $M=10^6\,\Msun$ and $\dotM=2000$ given a covering factor from 0.1 (left side) to 0.8 (right side).}
\label{fig:break}
\end{figure}

To interpret the observed Balmer break \citep{Ji2025,Naidu2025,Graaff2025b,Taylor2025b,Inayoshi2025b} and Balmer decrement \citep{Graaff2025,Chen2026a} seen in the spectrum of LRDs, 
we performed \textsc{Cloudy} calculations for a central AGN surrounded by a spherical envelope \citep[see also][]{Inayoshi2025b,Yan2026}, given an outflow model in \textsc{Cloudy}.
The \texttt{sphere} command is used to impose a closed geometry.
We assumed that the gas beyond the photosphere is illuminated and ionized by UV photons from a central source.
For the geometry in this study, UV photons can leak into the broad-line region (above the photosphere) through the low-density funnel, as $r_{\rm sc}$ is usually larger than $r_\ast$ by an order of magnitude.
The incident continuum is specified with the \texttt{AGN} command using the default SED parameters:
a disk temperature of $1.5\times10^5\,$K, a spectral index $\alpha_{\rm UV}=-0.5$ in UV and $\alpha_{\rm X}=-1$ in X-ray, and an X-ray-to-UV spectral slope $\alpha_{\rm OX}=-1.4$. 
The luminosity is normalized to the Eddington limit, i.e., close to $L_{\rm bb}$ (Eq.~\ref{eq:lum}).
A steady-state wind density structure is implemented with the \texttt{dlaw} command.
The dimensionless mass outflow rate is set as  $\dot{m} - 1$ as the model implies,
while the terminal velocity and inner radius are set by Equations (\ref{eq:vel}) and (\ref{eq:rstar}), respectively.
The calculation is stopped at a geometrical thickness $\Delta r=r_\ast$.
The gas metallicity is set to solar abundance, and the calculation is iterated to convergence.

We calculated the expected break strength by adopting a fiducial model with $M=10^6\,M_\odot$ and $\dot m=2000$, representative of the parameter space inferred above.
The Balmer break strength is defined as the ratio of the luminosity densities at 4000\,\AA\,to that of 3600\,\AA,
where the luminosity density is calculated as $F_{\lambda}=(1-{\rm CF})\cdot F_{\lambda,\rm inc}+{\rm CF}\cdot(F_{\lambda,\rm trans}+F_{\lambda,\rm dif})$,
where CF is the covering factor,
$F_{\lambda,\rm inc}$ is the incident continuum,
$F_{\lambda,\rm trans}$ is the transmitted  continuum,
and $F_{\lambda,\rm dif}$ is the diffused continuum.
The results assuming ${\rm CF}=0.1-0.8$ are shown in Figure~\ref{fig:break}, compared with observations quoted from \citet{Graaff2025}.
The model predicts a range of Balmer break strength generally consistent with observations.

\begin{figure}
\centering
\includegraphics[width=\linewidth]{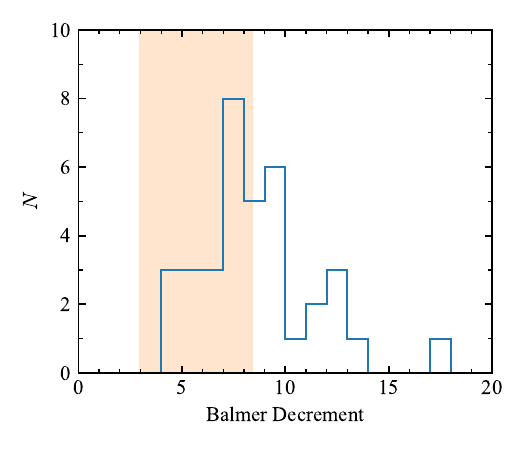}
\caption{Distribution of Balmer decrements in LRDs.
The shaded region shows the expected range from the outflow model with $M$ and $\dotM$ determined from the observed luminosity and temperature for each object.}
\label{fig:BD}
\end{figure}

Although Case B recombination predicts an intrinsic Balmer decrement about 3 under standard low-density nebular conditions \citep{Hummer1987}, 
this approximation may break down in a dense, partially ionized medium,
in which case resonant scattering and collisional excitation may suppress the emergent H$\beta$ relative to H$\alpha$ \citep{Chang2026} and lead to an enhanced Balmer decrement \citep{Nikopoulos2025,Yan2026}.
For each source, we inferred $M$ and $\dot{m}$ from the observed blackbody luminosity and temperature, and consequently the outflow properties. We then computed the corresponding Balmer decrement as $L_{\rm H\alpha}/L_{\rm H\beta}$,
where $L_{\rm H\alpha}$ and $L_{\rm H\beta}$ are the integrated H$\alpha$ and H$\beta$ luminosity, respectively.
Figure~\ref{fig:BD} compares the observed Balmer decrements with our model predictions.
The observed LRDs commonly exhibit H$\alpha$/H$\beta$ ratios of $\sim 4-17$ \citep{Graaff2025}, significantly larger than the Case-B value.
\textsc{Cloudy} calculations predict slightly enhanced Balmer decrements, matching the lower end of the observed distribution.

\section{Discussion}
\label{sect:discuss}

The outflow model driven by supercritical accretion, after matching the observed blackbody temperature and luminosity of the red continuum, can produce emission line FWHMs broadly consistent with the observed values, as well as Balmer break strengths consistent with the observed range.
The underprediction of the Balmer decrement may result from additional physics not included in the current model.
For example, the energetic outflow could drive fast radiative shocks and create an extended, shock-heated partially ionized zone, where collisionally excited H$\alpha$ would be heavily boosted relative to recombination-dominated H$\beta$, naturally yielding a steeper Balmer decrement.
In the following we discuss uncertainties of the model and possible caveats.

Numerical simulations \citep{Jiang2019, Sadowski2014,McKinney2014, Sadowski2015, Sadowski2016} suggest that, encircled by the optically thick outflow, there is a central low-density funnel where the radiation from the central accretion flow (e.g., a UV continuum) may escape and enter our line of sight after scattering \citep{Kawashima2009,Kawashima2012,Kitaki2017}.
If this funnel really exists, we would expect to see a population of AGNs with geometrically beamed emission at consistent redshifts of LRDs. 
Their fraction depends on the open angle of the funnel as well as the scattering optical depth in the funnel, which both affect the angular distribution of emission \citep{Zhang2005}.
At such high accretion rates, the low-density funnel is expected to be narrow \citep[with an open angle inversely proportional to the mass accretion rate; e.g.,][]{King2001}.
However, the high scattering optical depth in the funnel leads to a high-altitude last scattering surface \citep{Jiang2014}, causing central UV photons to escape into the scattering region since $r_{\rm sc} / r_\ast \approx 10$ in these cases.
The high density in the funnel is perhaps in line with the fact that LRDs are X-ray weak \citep{Maiolino2025}, because the X-ray corona cannot develop at high densities.
Anyway, the presence of the low-density funnel suggests that the covering factor should be less than unity. 
Recent spectroscopic observations of the $z\sim7$ LRD GlimmIr support this interpretation: flux variations in the continuum as well as the broad and narrow lines on a rest-frame timescale of $\sim 13$ days must allow direct lines of sight to the central engine \citep{Lambrides2026}. 
However, no numerical simulations have reached such a high accretion rate that allows us to estimate the open angle of low-density funnel and the angular distribution of UV emission through it; this should be studied in the future.

The mass derived from the outflow model is about one order of magnitude lower than those inferred using the single-epoch black hole mass estimator \citep{Maiolino2024}.
In the outflow model, the broad-line region constitutes an outflowing zone extending from the photosphere to the scattersphere. 
Since this region is not virialized, the virial mass of the black hole cannot be reliably derived from the broad-line width.

To sustain the high accretion rate, there should be a high-density reservoir near the center of the host galaxy.
The outflow will eventually interact with the surrounding gases, compress them, and produce shocks with relatively low velocities.
For example, shocks of $\sim 100\,\kms$ have been seen around many ULXs \citep{Pakull2002}. 
The shocked gas may produce narrow emission lines with a velocity dispersion depending on the shocked temperature, as well as blue shifted absorption features depending on the density.
This should also be investigated in future numerical simulations \citep{Huang2026}.

If LRDs have an effective duty cycle of order $0.1$, the corresponding cumulative active time is $\sim 10^8$ yr for a total cosmic time interval of $10^9$ yr.
In this model, however, the rate actually swallowed by the black hole is not $\dot{m}\sim 10^3$, but is regulated to be of order the Eddington rate, $\dot{m} \simeq 1$, while most of the excess supplied gas is expelled in the outflow.
The corresponding Salpeter timescale is $\sim 4.5\times10^7$ yr for $\eta=0.1$, so the black hole would grow only by a factor of $\sim \exp(10^8/4.5\times10^7)\sim 9$ over such an active period.
Therefore, the model does not necessarily require a very short-lived phase with catastrophic black hole growth.

The large mass supply rate mainly reflects the amount of gas processed through the inner region, rather than the net mass permanently accreted by the black hole.
In our picture, most of the excess gas is expelled as an optically thick outflow.
This material can interact with the surrounding dense gases and may partly cool, decelerate, and return toward the equatorial plane, thereby feeding the inflow again \citep{Kitaki2021}.
Thus, the cumulative processed mass over the LRD lifetime need not be interpreted as an equal amount of fresh gas that is consumed only once.
The instantaneous mass contained in the optically thick outflow is also modest.
A simple order-of-magnitude estimate of the instantaneous mass gives a characteristic scale of $\rho_i r_i^3 \sim 2.4\,M_\odot$ for our fiducial parameters.

We also note that in \cite{Kido2025}, a dense spherical envelope is required to confine the outflow and prevent it from disrupting the LRD.
In our picture, however, the envelope is not strictly spherical.
Instead, accretion inflows in the equatorial plane coexist with outflows launched between the polar and equatorial directions, as commonly seen in numerical simulations.
Therefore, the outflow does not necessarily quench the inflow, allowing accretion to be sustained.

\section{Conclusions}
\label{sect:conclusion}

Supercritical accretion onto black holes can drive an optically thick outflow, which can account for observed properties of LRDs:
\begin{itemize}[nosep]
    \item The observed blackbody luminosities and temperatures of LRDs correspond to $M=10^5-10^7\,M_\odot$ and $\dot m=1500-5000$ under this specific model.
    \item The FWHMs predicted by the outflow model that take into account both kinematic  broadening and scattering effects are broadly consistent with the observed broad-line widths, within a factor of 1.5 for more than 80\% of sources.
    \item The partially ionized hydrogen above the photosphere can produce Balmer break strengths comparable to those observed in LRD spectra.
    \item The predicted Balmer decrements are enhanced relative to Case B but remain below some observed values.
\end{itemize}

One caveat should also be noted.
Although a low-density funnel naturally implies a covering factor below unity, current simulations have not yet reached such extreme accretion rates considered here and therefore cannot reliably predict the funnel opening angle.

\begin{acknowledgments}
We are grateful to Yue Shen and the anonymous referee for constructive comments that improved the manuscript. 
We also thank Shuo Zhai, Jing Guo, and Zu Yan for helpful discussions regarding the use of \textsc{Cloudy}.
H.F. acknowledges funding support from the National Natural Science Foundation of China under the grant No. 12025301 and the Strategic Priority Research Program of the Chinese Academy of Sciences.
J.R.L. is supported by the China Postdoctoral Science Foundation under grant No. 2025M783227.
L.C.H. was supported by the China Manned Space Program (CMS-CSST-2025-A09) and the National Science Foundation of China (12233001).
\end{acknowledgments}

\bibliographystyle{aasjournal}
\bibliography{LRD}

\end{document}